\documentstyle[12pt,psfig,epsfig]{article}
\newcommand{\beq}{\begin{equation}}
\newcommand{\eeq}{\end{equation}}
\newcommand{\beqa}{\begin{eqnarray}}
\newcommand{\eeqa}{\end{eqnarray}}
\def\nue{{\nu_e}}

\def\numu{{\nu_{\mu}}}

\def\nutau{{\nu_{\tau}}}

\newcommand{\dm}{\mbox{$\Delta{m}^{2}$~}}

\newcommand{\cl}{\mbox{$^{37}{Cl}$~}}

\newcommand{\cnv}{\mbox{$\breve{\rm C}$erenkov~}}
\def\br{{$^{8}{B} ~$}}
\def\etal{{\it et al.}}

\newcommand{\ber}{\mbox{$^{7}{Be}$~}}
\begin{document}
\begin{center}
\bf{Solar Neutrino Experiments: An Overview}
\vskip 5pt
\end{center}
\begin{center}
Srubabati Goswami\footnote{email: sruba@mri.ernet.in}
\\
{\it Harish-Chandra Research Institute, Chhatnag Road, Jhusi,\\
Allahabad - 211-019, India}\\
\end{center}

\begin{center}
Abstract
\end{center}
This article describes the seven  experiments Homestake,
Kamiokande, SAGE, GALLEX, Super-Kamiokande, GNO and SNO which have
so far provided data on the measurement of the solar neutrino
fluxes. The detection mechanism, the salient features of the
detectors and the  results of each experiment are
presented. How the solar neutrino problem has evolved and became
more focused with each experimental data is summarized. The goals
for the future experiments are outlined.

\section{Introduction and History}
The sun is a copious source of electron neutrinos, produced in the
thermonuclear reactions that generate solar energy. The underlying
nuclear process is: \beq 4\;p \;\; \rightarrow \; \alpha \; + \;
2\;e^+\;+\;2\;\nu_e \;+\;25 \;{\rm MeV} \label{pp} \eeq About
97-98\% of the total energy released in the above process is in
the form of heat and light. The rest 2-3\% is carried away by the
neutrinos. The photons get scattered and re-scattered by
interaction with solar matter  and an average photon takes about
$10^{4}$ years to come out. Neutrinos are weakly interacting and
it takes about 8 minutes for them to reach to earth from sun. Thus
neutrinos carry information about the sun's interior.
However this very fact that the neutrinos are weakly interacting
makes it a difficult task to detect them and the typical
requirements are large detector volume, high detection sensitivity
and low background environments. Solar neutrinos were first
detected in 1968 by the pioneering $^{37}{Cl}$ experiment of
Raymond Davis in the Homestake gold mine in Lead, South Dakota
\cite{cl1}. The main motivation was to test the hypothesis of the
nuclear energy generation in stars. However when the results were
reported the observed flux was found to be less than the
theoretical prediction. This was the beginning of the solar
neutrino problem. It was also the first experimental signature of
the neutrino oscillation phenomena conjectured by Pontecorvo in
1957 and by Maki, Nakagawa and Sakata in 1962 \cite{bruno}. In
fact before the results from the \cl experiment came Pontecorvo
predicted that if solar neutrinos undergo  a change of flavour
then the observed fluxes can be less than the theoretical
expectations. For almost two decades \cl was the only experiment
measuring the solar neutrino flux . Next to join this pursuit was
the Kamiokande neutrino-electron scattering experiment which, not
only confirmed the deficit problem, but also verified that the
captured neutrinos are indeed of solar origin \cite{kam1}. A
further affirmation of the solar neutrino shortfall came from the
radiochemical $^{71}{Ga}$ experiments of the GALLEX \cite{gallex1}
and the SAGE \cite{sage1} collaborations. The triumph of the
Ga-experiments lies in the detection of the primary $pp$ neutrinos
thereby checking the basic hypothesis of stellar energy
generation. With all these experiments confirming the solar
neutrino problem it was realised that the solar neutrinos can be
used as an important tool to study neutrino properties and solar
neutrino research entered a new era of high statistics precision
experiments. Super-Kamiokande the upgraded version of the
Kamiokande experiment started operation in 1996 and declared its
first results in 1998 \cite{sksolar}. It not only confirmed the
solar neutrino deficit but it had enough statistics to divide the
events into energy and zenith angle bins enabling one to study the
incident neutrino spectrum shape and the presence of any
difference in the observed rate at day and at night. The most
recent results on this have come from the Sudbury Neutrino Observatory
in Canada which furnished direct evidence in favour of neutrino
flavour transitions \cite{snocc}. It also confirmed that the total
flux of the solar neutrinos, coming from \br decay,  
is in close agreement with the theoretical predictions
from solar model calculations \cite{snonc,snodn}. Finally on
December 6, 2002 the solar neutrino problem came to a full circle
with the reactor based experiment KamLAND in Japan providing
terrestrial evidence in favour of the Large Mixing Angle (LMA) solution 
based on the Mikheyev-Smirnov-Wolfenstein effect \cite{msw} 
to this \cite{kam_1}.

In the next section we discuss the production mechanism of the
solar neutrinos in the sun and the fluxes of these according to
Standard Solar Model(SSM) calculations. In the section 3 we describe the
seven solar neutrino experiments which have so far measured the
solar neutrino fluxes. In section 4 we discuss the solar neutrino
problem and how it has evolved as more and more data came. We also
briefly comment on the neutrino oscillation solution to the solar
neutrino problem. Finally we discuss the future experiments.

\section{Solar Neutrinos And The Standard Solar Model}
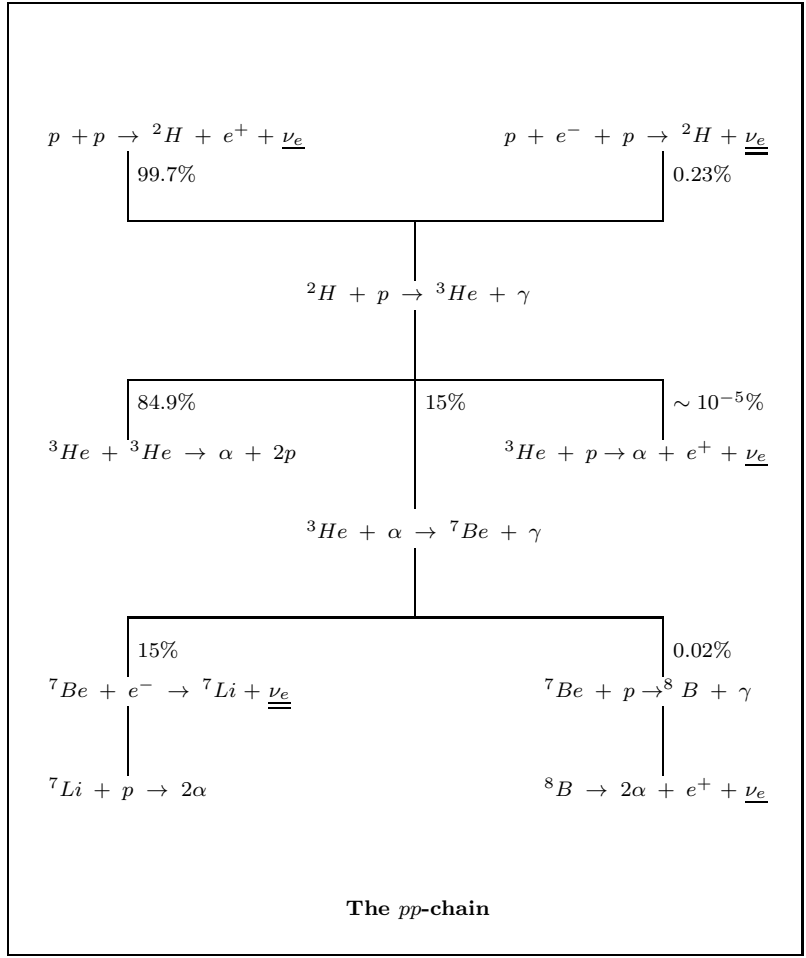
\begin{figure}
\setlength{\unitlength}{0.75pt}
\begin{picture}(450,600)(-50,150)
\begin{picture}(450,600)(0,0)

\scriptsize{
%\put(180,230){\normalsize Figure 1}
\put(0,270){\framebox(400,480)}
\put(20,680){$ p \; + p \; \rightarrow \; ^2H \; + \; e^+$ +
\underline {$\nu _e$}}

\put (250,680){$p \;+\;e^- \;+\;p \; \rightarrow \; ^2H$ +
\underline{\underline{$\nu _e$}}}

\put(60,675){\line(0,-1){35.0}}
\put(330,675){\line(0,-1){35.0}}

\put(65,660){99.7\%}
\put(335,660){0.23\%}

\put (60,640){\line(1,0){270.0}}
\put (205,640){\line(0,-1){30.0}}
\put(150,600){$^2H\;+\;p\; \rightarrow \;^3He\;+\;\gamma $}
\put (205,595){\line(0,-1){100.0}}
\put (60,560){\line(1,0){270.0}}
\put(60,560){\line(0,-1){30.0}}
\put(330,560){\line(0,-1){30.0}}

\put(65,545){84.9\%}
\put(210,545){15\%}
\put(335,545){$\sim 10^{-5}$\%}

\put(20,520){$^3He\;+\;^3He\;\rightarrow \;\alpha \;+\; 2p$}

\put(250,520){$^3He\;+\;p \rightarrow \alpha\;+\;e^+$ +
\underline{$\nu _e$}}

\put (205,475){\line(0,-1){35.0}}
\put (60,440){\line(1,0){270.0}}

\put(60,440){\line(0,-1){30.0}}
\put(330,440){\line(0,-1){30.0}}

\put(150,480){$^3He\;+\;\alpha \;\rightarrow \;^7Be \; + \;\gamma $}

\put(65,420){15\%}
\put(335,420){0.02\%}

\put(20,400){$^7Be\;+\;e^-\;\rightarrow \;^7Li$ +
\underline{\underline{$\nu _e$}}}
\put(270,400){$^7Be\;+\;p \rightarrow ^8B \;+\; \gamma$}

\put(60,395){\line(0,-1){35.0}}
\put(330,395){\line(0,-1){35.0}}

\put(20,350){$^7Li\;+\;p\;\rightarrow \;2 \alpha $}

\put(270,350){$^8B\; \rightarrow \;2 \alpha \;+\;e^+$ +
\underline{$\nu _e$}}

\put(170,290){{\bf The $pp$-chain}}}

\end{picture}
\end{picture}
\vskip -2cm \caption{The reactions of the $pp$-chain. The
probability of a particular reaction is shown as a percentage. The
neutrinos are shown underlined. Those with double underlines are
mono energetic.}
\label{ppchain}
\end{figure}
The main reaction for production of solar neutrinos is given by
eq. \ref{pp}. The reaction is the effective process driven by a cycle
of reactions ({\it e.g.} the $pp$-chain or the CNO cycle).  
The dominant set of reactions -- the $pp$-chain -- is
depicted in Fig. 3.1. 
These stellar thermonuclear fusion reactions, first understood by
Bethe in 1939, form the basis for the present solar models. The
algorithm of an SSM is to evolve a one $M_{\odot}$ homogeneous
cloud of hydrogen, helium and a small fraction of heavier elements
from about 4.6 billion years ago to match the current luminosity
and solar radius. These models are based on standard
thermodynamics and nuclear physics and assume that nothing happens
to the neutrinos once they are produced in the solar interior. 
The input parameters are the initial heavy element abundance,
radiative opacities, nuclear reaction rates, solar age, solar
luminosity etc. The predictions are the temperature, density and
composition profiles of the sun and the solar neutrino fluxes. The
most widely used  solar model  for neutrino flux calculations 
is the  standard solar model (SSM)
developed
by Bahcall and his collaborators \cite{bah,bp98,bp00}. However
there are a number of solar model calculations by different
groups \cite{turck}. All these show a remarkable agreement (to
within better than 10\%) between the predicted neutrino fluxes,
if same input parameters are employed. The solar model
calculations are also in excellent agreement with
helioseismological observations \cite{bp00}

\begin{figure}[htb]
%\vskip -0.5cm
%\begin{turn}{-90}
    \centerline{\psfig{file=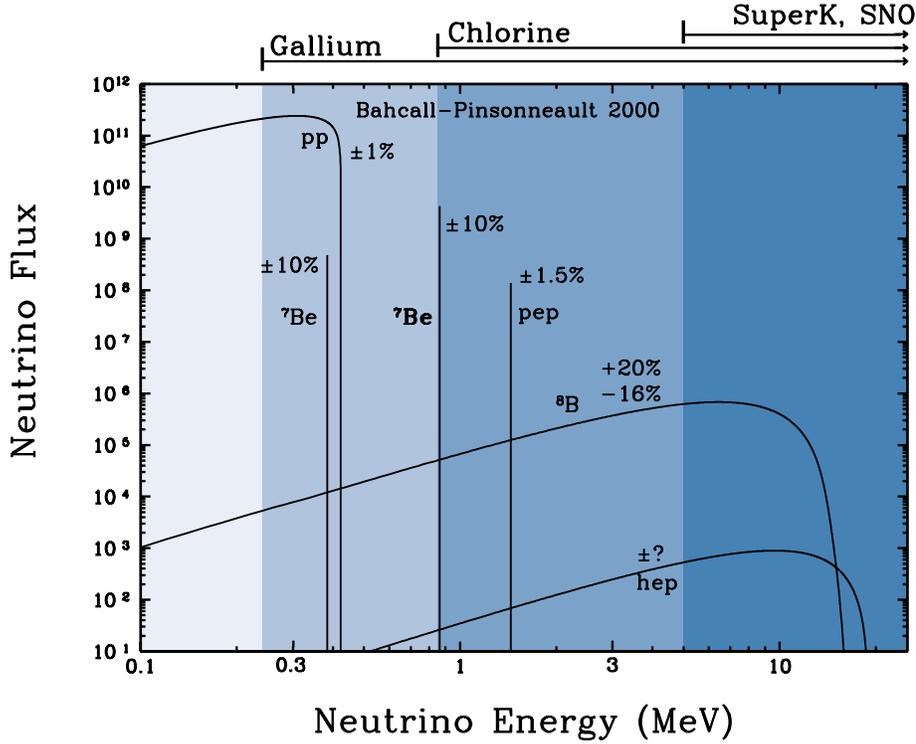,width=4.2in,angle=270}}
%\end{turn}
%\vskip -1cm
    \caption[The solar neutrino spectrum.]{\label{nuspec} Solar neutrino
spectrum in the standard solar model as a function of neutrino
energy. The continuous spectra are in units of
cm$^{-2}$MeV$^{-1}$s$^{-1}$, while the mono-energetic lines are in
cm$^{-2}$s$^{-1}$. Figure shows the $\pm 1\sigma$ uncertainties in
the model predictions of the various fluxes. Also shown are the
energy ranges over which the solar neutrino experiments are
sensitive \cite{jnb}.}
\end{figure}
Fig. \ref{nuspec} shows the solar neutrino spectrum from
\cite{jnb}. The SSM prediction of the pp($^8B$) solar neutrino fluxes
are least (most) uncertain.
These uncertainties 
are  associated with the input parameters and considerable effort
is directed towards reducing the margin of these uncertainties.

\section{Solar Neutrino Experiments}

The present solar neutrino detectors are of two types
{\footnote {For a discussion on other type of detectors like cryogenic and bolometric detectors and 
geochemical solar neutrino experiments see \cite{kirsten}.} \\
(i){Radiochemical detectors}\\
(ii){\cnv detectors}

In radiochemical experiments one employs a nuclear reaction of the
form \beq ^A{Z}+\nue \rightleftharpoons ^A{(Z+1)} + e^- \eeq where A
denotes the mass number and Z  denotes the proton number of the
nuclei. The Q value of the reaction determines the threshold
neutrino energy for the detector. The target is exposed to the sun
for a certain period of time after which the product is extracted
using radiochemical techniques and counted by electron capture decays of the 
product nucleus. Thus it is not possible to determine
the neutrino energy and direction.
The experimentally challenging concept involved in this technique was the 
chemical separation of a small number of product nuclei from a large mass 
of target atom and the past and present radiochemical experiments 
have unequivocally demonstrated the validity of this procedure. 
The main systematic uncertainties involved in radiochemical experiments 
are due to extraction efficiency, counting efficency and background. 

In a \cnv detector on the other hand the charged leptons produced from
neutrino interactions produce
\cnv light while passing through the detector. This is
recorded by photomultiplier tubes
which produce an electrical pulse that is registered by the data
acquisition electronics.
Since the signal to noise ratio is very small at lower energies such
a detector usually has higher threshold energies.
But it has important advantages\\
(i)it is a detector in real time;\\
(ii) it gives directional information because
the \cnv light allows a reconstruction of the incoming neutrino
track; \\
(iii)the recoil electron energy distribution gives information about
the incident neutrino energy spectrum. 

\subsection{Radiochemical Detectors}

\subsubsection
{The Homestake Experiment}

%\underline{The Reaction}\\
The reaction involved in the detection is
\beq
\nu_e + ^{37}{Cl} \rightarrow ^{37}{Ar} + e^{-}.
\label{37cl}
\eeq
which has a threshold of 0.814 MeV and is sensitive to
the $^{8}{B}$ and $^{7}{Be}$ neutrinos.
This method was first suggested by Pontecorvo in 1946 and
independently by L. Alvarez in 1949.
In 1959 it was realised that the measured
 $^3{He}(^3{He},2p)^4{He}$ cross-section
is more than expected on the basis of theoretical calculations
\cite{holm}. This signified that the pp chain does not terminate
with the reaction $^3{He}(^3{He},2p)^4{He}$ continues further
resulting in the production of $^7{Be} $ and $^8B$ neutrinos as
shown in fig. \ref{ppchain}. Both of these have energies above
0.814 MeV, the threshold for reaction \ref{37cl}. This
encouraged Raymond Davis to 
construct the first detector for solar neutrinos  
in the Homestake gold mine in South Dakota using
$^{37}{Cl}$ in the form of perchloroethylene ($C_2Cl_4$). The plus
points were the following
\\ (i) $C_2Cl_4$, a liquid at room temperature, was inexpensive
and easily available; \\ (ii)the daughter nucleus $^{37}{Ar}$
being an inert gas could be extracted easily;
\\
(iii)
 the  transition of the $^{37}{Cl}$ to the
isobaric analogue state in $^{37}{Ar}$ for energies greater than
5.8 MeV enhances the neutrino capture cross-section considerably
making this process particularly suitable for studying the high
energy $^8B$ spectrum \cite{iso}. 

%\underline{The Detector}\\
The detector consists of a cylindrical tank 6.1m in diameter and
14.6m long. About 95\% of the detector volume is filled with 133
tons of perchloroethelene. The underground location of 4200 m
underground cuts down the production of $^{37}{Ar}$ in the
detector due to cosmic rays. The $^{37}{Ar}$ produced is separated
chemically from the ${C_2}{Cl_4}$, purified and counted in low
background proportional counters. The counting depends upon
observing the 2.82 KeV Auger electrons from the $e^{-}$ capture
decay of $^{37}{Ar}$ (half life = 35days). 

%\underline{Results}\\
The Homestake detector started operating from 1968 and has reported data
taken in 108 runs over the period 1970-1994.
The third column of Table \ref{bp00tab1}
shows the predictions \cite{bp00} for the neutrino capture
rates in the Cl experiment for the different neutrino sources.
Also shown for the Cl detector are the total predicted rate and
the $\pm 1 \sigma$ uncertainties
in the model calculations. The numbers quoted are in a convenient unit
called SNU, defined as,
%\be
$1 ~{\rm SNU} = 10^{-36} {\rm events/target~atom/second}$.
%\ee
The observed rate of solar neutrinos in the experiment is
\cite{cl}  
%\beq 
%{\rm Observed~Rate} = 
$2.56 \pm
0.16 {(\rm stat)} + 0.16 {(\rm syst)}$ {\rm SNU}.  
%\eeq 
Compared to
the prediction of Table 
\ref{bp00tab1}, this gives a ratio of observed to expected SSM
rate of $0.335 \pm 0.029$ implying a deficit in the solar neutrino flux.
No significant time variation was reported by the data.
The sources of systematic uncertainties
are extraction efficiency, counting efficiency, neutrino
production due to non solar sources, event selection and variation
in the half-life of the decaying background component 
resulting in a systematic uncertainty of $\sim 7\%$ in a single
run \cite{cl}. 
Since very few $^{37}{Ar}$ atoms are produced in each run the
statistical uncertainty is 30-50\%. However the cumulative
statistical uncertainty for the total 108 runs is comparable to
the systematic uncertainty. The main uncertainty in the
theoretical prediction is due to the uncertainties associated with
the $^8B$ flux calculation.

\subsubsection{The $^{71}{Ga}$ Experiments}
There are three
experiments SAGE, GALLEX and GNO  which
employ the following reaction (suggested first by Kouzmine in
1965): \beq \nu _e \; + \; ^{71}Ga \; \rightarrow \;^{71}Ge\; +
\;e^- \label{gaeq} \eeq This reaction has a low threshold of 0.233
MeV and the detectors are sensitive to the basic $pp$ neutrinos.
Since the $pp$-chain is mainly responsible for the heat and light
generation in the sun, detection of these neutrinos constitute an
important step towards confirming the accepted ideas of
solar energy synthesis. Also the predicted flux is relatively free
of the astrophysical uncertainties.  The SSM prediction for this
according to ref. \cite{bp00} is 130$^{+9}_{-7}$ SNU. In Table
\ref{bp00tab1} we present the contribution of each of the sources 
as well as the total rate according to \cite{bp00}. 

\begin{table}[htbp]
    \begin{center}
        \begin{tabular}{||c||c|c|c||} \hline\hline
         {\rule[-3mm]{0mm}{8mm}
         source} & Flux & Cl & Ga \\
    [1ex]     & ($10^{10}$ cm$^{-2}$s$^{-1}$) & (SNU) & (SNU)\\
    [1ex]\hline\hline
         {\rule[-3mm]{0mm}{8mm}
         $pp$}  &  5.95($1.00^{+0.01}_{-0.01}$)
          &  0.0 & 69.7 \\[1ex]
         $pep$ &  1.40$\times 10^{-2}$($1.00^{+0.015}_{-0.015}$)
          & 0.22 & 2.8 \\[1ex]
         $hep$ &  9.3$\times 10^{-7}$ & 0.04 & 0.1\\[1ex]
         $^{7}Be$ & 4.77$\times 10^{-1}$($1.00^{+0.10}_{-0.10}$)
          & 1.15 & 34.2 \\[1ex]
         $^8B$ & 5.05$\times 10^{-4}$($1.00^{+0.20}_{-0.16}$)
          & 5.76 & 12.1 \\[1ex]
         $^{13}N$ & 5.48$\times 10^{-2}$($1.00^{+0.21}_{-0.17}$)
          & 0.09 & 3.4 \\[1ex]
         $^{15}O$ & 4.80$\times 10^{-2}$($1.00^{+0.25}_{-0.19}$)
          & 0.33 & 5.5 \\[1ex]
         $^{17}F$ & 5.63$\times 10^{-4}$($1.00^{+0.25}_{-0.25}$)
          & 0.0 & 0.1 \\ [1ex]\hline\hline
            &  & & \\
         \raisebox{1.5ex}[0pt] {Total} &  &
         \raisebox{1.5ex}[0pt] {$7.6_{-1.1}^{+1.3}$}&
         \raisebox{1.5ex}[0pt] { $128_{-7}^{+9}$} \\
         \hline\hline

         \end{tabular}
      \end{center}
\vskip -0.5cm
      \caption[The BPB00 predictions]{\label{bp00tab1}
    The predictions for the solar neutrinos
    fluxes and neutrino capture rates in the Cl and Ga detectors
    from \cite{bp00}.
    The expected \br flux is
    5.05$\times 10^{6}{\rm cm}^{-2}{\rm s}^{-1}$.}
\end{table}
\vskip 10pt

\noindent {\bf {$\bullet$
{Soviet American Gallium Experiment(
SAGE)}}}\\
The SAGE experiment in the deep underground  Baksan Neutrino
Observatory has been measuring the solar neutrino capture rate
since 1990 and it is still continuing to take data. It uses
50 tons of metallic $Ga$ in liquid form as the target. This is
contained in seven chemical reactors. The advantage of the
metallic target is its low sensitivity to radioactivity compared
to any other form. The $^{71}{Ge}$ produced is separated by
stirring vigorously with a mixture of hydrogen peroxide and dilute
hydrochloric acid. The $Ge$ is extracted as $GeCl_{4}$ which is
subsequently converted to $GeH_{4}$ by sodium borohydride. The
counting is done in a proportional counter by observing the Auger
electrons and X-rays emitted in the $^{71}{Ge}$ electron capture
decay producing an L peak at 1.2 KeV and K peak at 10.4 KeV. data
of 92 runs during the period from January 1990 to December 2001
gives a solar neutrino capture rate of \cite{sage}
%\begin{center}
$%{\rm Observed~Rate} = 
70.8~ \pm~^{5.3}_{5.2}~(\rm stat)~^{+5}_{-7}~(\rm syst)
~SNU$.
%\end{center}
SAGE has performed a  calibration test with a $\sim$ 0.5 MCi
$^{51}{Cr}$ source. This gave results consistent with expectations
demonstrating that there are no unknown experimental
errors that can count for the observed deficit.

\vskip 10pt
\noindent 
{\bf{
$\bullet$
GALLEX}}\\
The GALLEX experiment is located in the Gran Sasso underground
laboratory in Italy. The detector consists of 30 tons of $Ga$ in the
form of $GaCl_{3} - HCl$ solution. $Ge$ is produced in the form of
volatile $GeCl_{4}$. After an exposure of about three weeks the
$GeCl_{4}$ formed is extracted by bubbling nitrogen through the
solution and then passing through two gas scrubbers where the
$GeCl_{4}$ is absorbed in water. This is finally converted to the
counting gas $GeH_{4}$ (germane).
The counting is done as in SAGE.
The chemical form of Ga used in GALLEX is advantageous for the 
extraction of the product. 

GALLEX has taken data during the period May 1991 to January 1997
and the combined result of a total of 65 runs is
%\begin{center}
$
%{\rm Observed~Rate} = 
77.5~ \pm~ ^{+7.6}_{-7.8}~
~SNU$, 
%\end{center}
where the statistical and systematic errors are combined in
quadrature \cite{gallex}. GALLEX has performed an overall calibration of their
detector using the two neutrino lines at 746 KeV (90\%) and 426
KeV (10\%) of a $^{51}{Cr}$ source. Following the same extraction
and counting procedures as in the solar neutrino runs, the ratio
of measured $^{71}{Ge}$ to the expected value obtained is
$R~=~1.04 \pm 0.12$, {\it viz.} in good agreement with the
expectations.

\vskip 10pt
\noindent
{\bf{
$\bullet$ Gallium Neutrino Observatory (GNO)}}

GNO is the upgraded version of GALLEX. The target and extraction procedure
used in this experiment is the same as GALLEX but the counting system was
modified resulting in an improvement in the noise to background ratio.
This is taking data since 1998.
From a total of 35 solar runs GNO reports an observed rate of
67.7 $\pm$ 7.2(stat) $\pm$ 3.2(syst) SNU \cite{gno}.
GNO is still continuing to take data. The future  goals are \\
(i)to reduce the systematic error at 3\% level; \\ (ii)  to use a
more sophisticated analysis based on neural networks to re-analyze
the data;
\\
(iii)to improve the knowledge of the $\nue$ capture cross section of
$^{71}{Ga}$  to better than 5\% using a $>$ 2.5 MCi $Cr$ source.

\subsubsection{The Kamiokande Experiment}
The Kamiokande  (Kamioka Nucleon Decay Experiment) experiment
\cite{kam} was started in 1983 to look for proton decay.
In 1985 it was renovated
for the solar neutrino research to test the anomaly observed by
the \cl experiment by direct detection techniques.
The Kamiokande
detector, located in a deep mine at Kamioka, Japan,
uses 4500 ton of pure water of which the inner 680 ton was
employed as the fiducial volume for solar neutrino observations.
Both the outer and the inner walls of the cylindrical detector
are lined with Photomultiplier tubes (PMT).
These
detect  the \cnv
light emitted by electrons which are scattered in the forward
direction by solar neutrinos \beq \nu_e~+ ~e \rightarrow
~\nu_e~+~e. \label{nuescatt} \eeq Unlike eq. (\ref{37cl}),
neutrinos and antineutrinos of all flavor can contribute to the
above. The $\numu$  and $\nutau$ react {\it via} the neutral
current which is suppressed by a factor of 1/6 compared to the
$\nu_e$ interaction which can be mediated by both charged and
neutral currents.
The recoil electron energy threshold in
Kamiokande
is 7.5 MeV. Thus it can sample only the $^{8}{B}$ neutrinos.
Using the timing information and the ring pattern of the hit 
PMTs the vertex position and the direction of the recoil electrons 
are reconstructed. These can be used to 
provide information on
the incoming neutrino direction. Kamiokande  
found an excess of events peaking in the direction of the
sun and thus verified for the first time that the captured neutrinos 
originate from the sun. 
The observed rate is 
%\begin{center}
%Observed Rate = 
%(
2.82 $\pm 0.19(stat) \pm 0.33(syst)) \times
10^{6}~cm^{-2}s^{-1}$ \cite{kam}
%\end{center}
while the theoretical prediction according to \cite{bp00} is
$5.05 \times 10^6~cm^{-2}s^{-1}$. The recoil spectrum
measured by Kamiokande is consistent with the solar $^8{B}$
neutrino spectrum, with an overall reduction in flux.

\subsubsection{SuperKamiokande}
SuperKamiokande is an upgraded version of Kamiokande with the detector 
volume
increased to 50 ktons 
of which the fiducial volume is 22.5 kilotons.
It is viewed by $\approx$ 13,000 PMTs. 
It has collected data for the period between May 31, 1996 to 
15 July 2001 (1496 days). 
In July 2001 data taking was stopped for detector upgrade. In
November 2001 there was an accident in the SK detector which
destroyed 6777 inner and 1110 outer photomultiplier tubes
respectively. Partial reconstruction of the detector has been
achieved till date.
During the 1496 effective days SK has observed 22400 $\pm$ 800 
solar neutrino events. 
The total observed rate in SK
is \cite{smy2002} 
%\beq
%Observed Rate = 
$2.35 \pm 0.02~(\rm stat) \pm 0.08~(\rm syst) 
\times 10^6/cm^2/sec$
%\eeq
almost half the prediction of the SSM of \cite{bp00}.
The threshold energy for recoil electrons in SK was 
6.5 MeV to start with and then it was
brought down to 
5.0 MeV.  
%For energies $>$ 15 MeV the $hep$ neutrino flux becomes dominant. 
%Taking into account the 
%energy resolution SK considered an optimum range from 18.0 to 
%20 MeV to look for the hep neutrinos. They found $4.9 \pm 2.7$ 
%events limiting the $hep$ flux to be 73.0 $\times 10^3/cm^2/sec$ at 90\%
%C.L. almost 8 times less than the SSM flux of Table 1. 

SK has sufficient statistics to divide both the day time and night time 
observed  events 
into recoil electron energy bins. 
Such a precision measurement would require an accurate calibration of the 
energy scale  which is performed using 
an electron linear accelerator.  
Fig. \ref{dnspec} plots data/SSM as a function of the recoil electron 
energy. 
The plot does not reveal any significant spectral distortion. 
The correlated systematic uncertainties shown by gray lines are  
due to the calculation of the $^8B$ neutrino spectrum, the absolute 
energy calibration and the energy resolution. 
The fig. \ref{dnspec} also shows the 
the day/night asymmetry as a function of energy. 
This asymmetry is seen to be consistent with zero at all energies. 

\begin{figure}[htb]
%\vskip -4.5cm
    \centerline{\epsfig{file=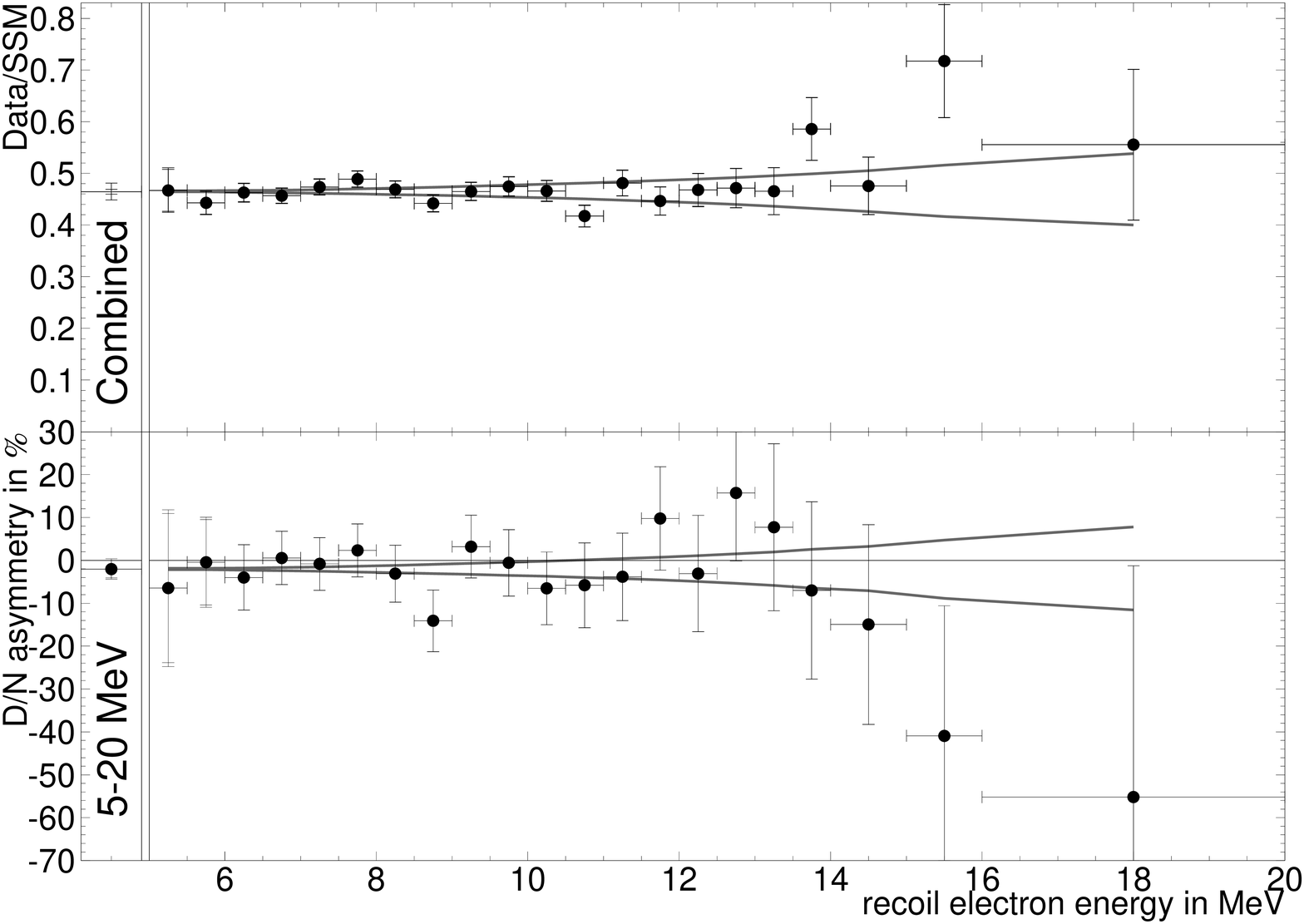,width=3.5in,height=4.0in}}
\vskip -0.5cm
    \caption[Super-K spectra at day and night.]{\label{dnspec}
SK recoil energy spectra along with the $\pm 1\sigma$ error bars
for 1496 day data. 
Also shown is the variation of the day/night asymmetry with energy
from \cite{smy2002}.}
\end{figure}

In order to study the spectral distortion and the 
day-night variation of the solar neutrino flux 
in greater details, the SK
collaboration have divided their data 
into eight energy bins and seven solar zenith angle bins 
\cite{smy2002}. 
The data shows no significant distortion of the $^8B$ spectrum 
with solar zenith angle variation. 
This binning to study the energy dependence of the
suppression rate as well as the predicted day night asymmetry
together in the most efficient manner. Apart from the zenith angle
dependence SK has also reported the seasonal variation of the
solar flux. The data is consistent with the expected annual
variation due to orbital eccentricity of the Earth.

%%%%%%%%%%%%%%%%%%%%%%%%%%%%%%%%%%%%%%%%%%%%%%%%%%
\subsubsection {The Sudbury Neutrino Observatory (SNO)}
%%%%%%%%%%%%%%%%%%%%%%%%%%%%%%%%%%%%%%%%%%%%%%%%%%

%\underline{The SNO Detector}

SNO is an imaging \cnv heavy water detector
containing 1 kton of pure ${\rm D_2O}$. It is 
located at a depth of 2092m (6010m water equivalent) in
Creighton mine near Sudbury, Ontario, Canada. The heavy water is
contained in an acrylic vessel 12m in diameter and thickness 5.5
cm. A geodesic support structure mounted with an array of 9456
photomultiplier tubes surrounds the acrylic vessel. The detector
is immersed in 7 kton of ultra-pure $H{_2}O$ within a barrel
shaped cavity.\\
The deuterium in the heavy water enables the detection of
all three types of neutrinos.
There are three main detection processes
\\
\beq \nu_e + d \rightarrow p + p +e^- ~~~~(CC) \label{nued} \eeq
\beq {\nu_x} + e^- \rightarrow {\nu_x} +e^- ~~~~(ES) \eeq \beq
\nu_x + d \rightarrow n + p + \nu_x~~~~~(NC) \label{neutral} \eeq
The ES reaction  is the same as used by SK. Its great
advantage is the strong directional correlation with the sun. For
the CC reaction the energy of the recoil electron is strongly
correlated with the incident neutrino energy and thus can provide
information on the $^8{B}$ energy spectrum. It also has an angular
correlation with the sun which goes as $1 - 0.34 cos
\theta_{\odot}$. It has a much larger cross section than the ES
reaction and can produce more statistics. 
Both the ES and
the CC reaction are detected by the \cnv light produced by the recoil
electron. The threshold kinetic energy of the recoil electron is 5
MeV for SNO ES and CC. For the NC reaction the threshold neutrino
energy is 2.2 MeV, the binding energy of the deuteron. The
produced neutron is thermalised and can be captured by another
nucleus. This nucleus emits a gamma ray which Compton scatters
electrons which are observed through \cnv radiation. The detection
efficiency depends on the neutron capture efficiency. Neutrons can
be captured directly on deuteron to give tritium and a 6.25 MeV
gamma which carries the tritium binding energy. 
But
this process is not very efficient with a capture efficiency of
29.9\%. The neutron capture efficiency can be increased to about
83\% by adding salt (NaCl) in heavy water. The Chlorine has a high
capture cross-section for the neutrons. This process results in a
gamma ray cascade with a peak around 8 MeV. 
Since
in both the processes the neutron is eventually observed through
the \cnv light generated by the electron the NC signal is
entangled with the CC and ES signal. To observe exclusively the
neutrons the SNO collaboration will employ $^3{He}$ proportional
counters (neutral current detectors) which will be installed in
the heavy water. $^3{He}$ has a large capture cross sections for
the thermalised neutrons producing an energetic proton-triton pair
which ionise the gas in the counter resulting in an electrical
pulse. Although this method requires complex hardware the
advantage is that it gives an independent measurement for NC
events.
\begin{figure}[t]
\centerline{\epsfig{figure=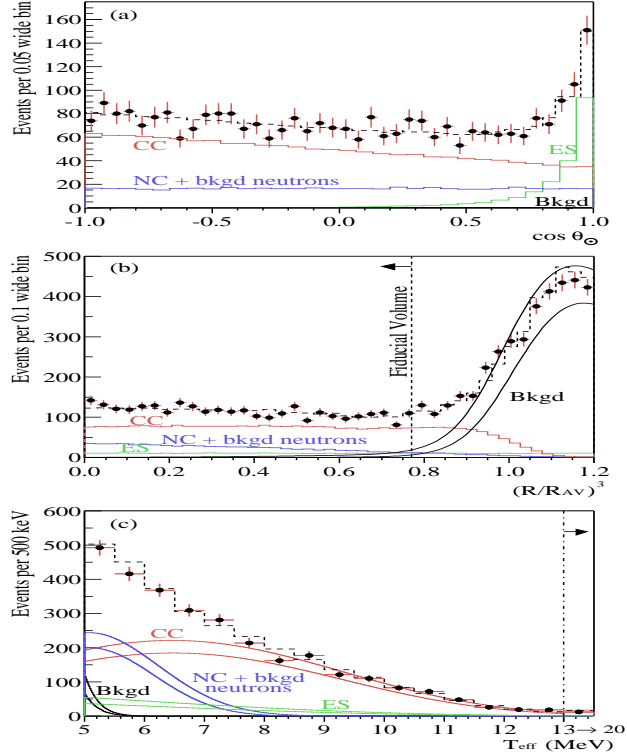,height=4.in,width=3.5in}}
\caption{No. of events observed at SNO
(a)vs. $\cos\theta_{odot}$,(b)$(R/R_AV)^3$,(c)kinetic energy.
Also shown are the Monte Carlo predictions for the
CC, ES and NC+background events. The dashed line represents the
summed components. The uncertainties are $\pm 1 \sigma$.
The figure is taken from from \cite{snonc}.
}
\label{snospec}
\end{figure}

%\underline{Results}
The unique feature of
SNO is the NC reaction which is equally sensitive to all neutrino
flavours. Thus this can provide a direct model independent
measurement of the $^{8}{B}$ flux. The comparison of the CC and NC
rate can provide the "smoking gun" evidence in favour of a
$\nu_\mu/\nu_\tau$ component in the sun. This evidence can also be
obtained to some extent by comparing the CC and ES rates but the
later has a limited sensitivity to NC events.

SNO has declared the data taken between the period
Nov 2, 1999 to May 28, 2001 consisting of 306.4 live days in April 2002 \cite
{snonc}. The neutral current data corresponds to those due to neutron 
capture on
deuteron.  
The fig. \ref{snospec} shows the no of events vs
the kinetic energy of the recoil electrons
$T_{eff}$,
 the reconstructed direction of the events w.r.t the sun
$\cos\theta_{odot}$ and the volume weighted radial variable $(R/R_{av})^3$.
Since all the three processes are observed through the \cnv light produced by electrons the SNO collaboration gives the total number of the CC+ES+NC events. 
To determine the individual rates one needs a further assumption 
of standard 
spectral shape above $T_{eff} = 5$ MeV.  
The flux of $^8{B}$ neutrinos measured in each reaction in SNO assuming no
energy distortion are \\
\beq
{\phi_{CC}} = {1.76 \pm 0.06(stat) ^{+0.09}
_{-0.11}(sys) \times 10^6 cm^{-2} s^{-1}}
\eeq
\beq
{\phi_{ES}}= {2.39 \pm 0.24 (stat)^{+0.12}_{-0.12}(sys)
\times 10^6 cm^{-2} s^{-1}}
\label{phies}
\eeq
\beq
{\phi_{NC}} = {5.09 \pm 0.44(stat) ^{+0.46}
_{-0.43}(sys) \times 10^6 cm^{-2} s^{-1}}
\label{phinc}
\eeq

SNO has also measured the spectra and rates at day and night. 
For the charged current events, assuming a constant spectral shape
SNO finds a day-night flux difference  
of 14\% $\pm6.3\%^{+1.5}_{-1.4}\%$ of the average rate \cite{snodn}. 
The dominant source of the systematic uncertainties in SNO are due to the 
energy calibration and resolution.   

\section{Summary of the experimental results and their implications}

In Table \ref{rates} we present the ratio
of the observed to the expected total rates
in the
Ga, Cl, SK and SNO experiments.

\begin{table}[htbp]
    \begin{center}
\begin{tabular}{||c|c|c||} \hline\hline
{\rule[-3mm]{0mm}{8mm}
experiment} & $\frac{obsvd}{BPB00}$ & composition \\ \hline \hline
Cl &
0.335 $\pm$ 0.029 & \br (75\%), \ber (15\%)
\\\hline
Ga & 0.584 $\pm$ 0.039 &$pp$ (55\%), \ber (25\%), \br (10\%)
\\\hline
SK & 0.459 $\pm$ 0.017  &
\br (100\%)
\\ \hline
SNO(CC) & 0.347 $\pm$ 0.027 &  \br (100\%)\\\hline
SNO(ES)& $0.473 \pm 0.074$ & 
\br (100\%)\\ \hline
SNO(NC) & 1.008 $\pm$ 0.122 & \br(100\%) \\
\hline\hline
\end{tabular}
      \end{center}
\vskip -0.3in
 \caption[The observed solar neutrino rates]{\label{rates}
   The ratio of the
observed solar neutrino rates to the corresponding SSM predictions
of \cite{bp00}. 
The Ga rate corresponds to the combined
SAGE and GALLEX+GNO data. Also shown is the composition of the
observed fluxes.  The SNO rates are obtained assuming no spectral 
distortion}
\end{table}

The observed $\nu_e$ 
fluxes in all the experiments are less than the expectations from
SSM and this constitutes the essence of the solar neutrino problem.
However as more and more data accumulated it  became more focused.
If we combine the $^{8}{B}$ flux observed in SK with the Cl
experimental rate, it shows a strong suppression of the $^{7}{Be}$
neutrinos. The pp flux constrained by solar luminosity along with
the $^{8}{B}$ flux observed in SK leaves no room for the
$^{7}{Be}$ neutrinos in Ga. This vanishing of the $^{7}{Be}$
neutrinos renders a purely astrophysical solution to the solar
neutrino problem impossible and neutrino flavour conversion was
conjectured as a plausible solution. This was beautifully
confirmed by the SNO data.

Since the CC reaction is sensitive only to $\nu_e$ and the ES
reaction is sensitive to both $\nu_e$ and $\nu_\mu/\nu_\tau$ a
higher ES flux would signify the presence of $\nu_\mu/\nu_\tau$.
The combination of SNO CC and SK ES data provides a 3.3$\sigma$
signal for $\nu_e$ transition to an active flavour (or against
$\nu_e$ transition to solely a sterile state).
The combination of the SNO NC and CC data 
gives 
\beq
\phi_{\mu\tau} = 3.41 \pm^{+0.66}_{-0.64} \times 10^6 cm^{-2} s^{-1}
\eeq
This
establishes
neutrino flavour conversion to a
state containing an
active neutrino component at 5.3$\sigma$ level.
Incorporating the SK measurement as an extra constraint 
confirms 
this at 5.5$\sigma$ level. 

There can be various mechanisms of neutrino flavour conversion
among which most popular is neutrino oscillations which requires
neutrinos to have mass and mixing. If one takes all the data from
the seven experiments into account namely the data on total rates
from Cl and Ga, the SK zenith-angle spectrum data and the SNO
spectrum data and perform a global $\chi^2$ -analysis two regions
in the mass squared difference ($\Delta m^2$) and mixing angle
(expressed as $\tan^2\theta$) parameter space  remain allowed. The
favoured solution from the solar data is the LMA region which
covers a range $3\times 10^{-5}$ eV$^2$ $ \leq \Delta m^2 \leq
3\times 10^{-4}$ eV$^2$ and {$0.25 \leq \tan^2\theta \leq 0.87$}
at 99.73\% C.L. ($3\sigma$) \cite{assd}. 
Confirmation in favour of this solution came recently
from the KamLAND experiment in Japan which used  reactor
$\bar{\nu_e}$s to look for neutrino oscillation \cite{kam_1}.
The first KamLAND results have already demonstrated remarkable 
capability to constrain the LMA region from its data 
on the spectrum of $\bar{\nu_e}$. At 99\% C.L. the LMA region 
gets bifurcated into two zones -- 
 a low $\Delta m^2$ region (low-LMA)  around the global best-fit
point with $\dm=7.17\times 10^{-5}$ eV$^2$ and $\tan^2\theta_{12}=0.44 
$and
a high $\Delta m^2$ region (high-LMA) around
$\dm=1.49\times 10^{-4}$ eV$^2$ 
and $\tan^2\theta_{12}=0.43$
having a less
(by $\approx 2\sigma$) statistical
significance \cite{kl}.

\section{Future Prospects}
With KamLAND confirming the LMA solution and the SK and SNO 
experiment measuring the solar $^8B$ flux  the two major goals 
of solar neutrino research are
\\
(i)precise determination of the neutrino mass and mixing parameters
\\
(ii)to observe the low energy end of the solar neutrino spectrum 
consisting of the pp,CNO and the $^7{Be}$ line. 

The Borexino liquid scintillator detector located in the
Gran-Sasso underground laboratory in Italy will measure the $^7Be$
flux via neutrino electron scattering reaction \cite{borexino}. The main problem
in the measurement of the low energy fluxes is the background
reduction. A 4 tons prototype called counting test facility (CTF)
has demonstrated extremely low radioactive level ($10^{-16}$g/g of
U/th) in Borexino. With KamLAND confirming LMA solution to the solar neutrino
problem Borexino should observe a rate of $\sim 0.64$ and no day
night asymmetry. 

Till now the $pp$ neutrinos have been observed in the Ga experiments 
using radiochemical techniques. 
Among these SAGE and GNO will continue taking
data and accumulate statistics.  
There is
continued exploration of a 100 ton Ga experiment for reducing the
statistical error. Ga experiments with increased statistics and 
reduced systematics can determine the pp flux more accurately. 
A radiochemical Li detector is proposed to measure the 
CNO fluxes \cite{li}. 
But the crying need in the field of low energy solar neutrinos is 
real time detectors.
The daunting task at these low energies is to reduce the 
radioactive backgrounds and various 
techniques are being discussed.
XMASS is a liquid
xenon scintillator detector planned to be installed in Kamiokande
site. It will use the neutrino electron elastic scattering
reaction to measure the pp flux \cite{scho}. 
Other real time experiments which will use $\nu-e$ scattering   
for measuring the pp and CNO fluxes
are HELLAZ, HERON,CLEAN,MUNU and GENIUS projects. 
HELLAZ is studying the possibility of measuring electron tracks
generated in  pressurised He while HERON is planning for bolometric detection 
using super-fluid He \cite{hellaz,heron}. 
MUNU will use projection chamber filled with $CF_{4}$ \cite{munu}. 
GENIUS is a proposed double beta decay experiment  having simultaneously 
the capability of real time detection of low energy neutrinos 
by suitably reducing the background \cite{genius}.
Charged current reactions for measuring low energy neutrino fluxes
are studied  
in LENS, MOON and SIREN proposals \cite{scho}. 
The measurement of low energy fluxes are technically quite challenging and
it is still at a nascent stage with feasibilty studies of the various 
proposed projects underway. 

\section{Conclusions}
Solar neutrino experiments have played a key part in
determining the fundamental properties of neutrinos.  
They also provide sensitive test of solar models.  
Apart from establishing the
presence of $\nu_\mu/\nu_\tau$ component in the solar $\nu_e$ flux
with a significance of  $5.3 \sigma$ the data
released by the SNO collaboration 
demonstrated that the total $^8B$ flux measured in the SNO NC
reaction is in close agreement with the Standard Solar Model
predictions. The first and the second generation of experiments
have contributed a great deal in advancement of our knowledge in
the fields of particle physics, astrophysics and experimental
techniques. 
Research and development studies are  in progress for
new experiments to measure the low energy  solar
neutrino fluxes for realising the goal of 
performing solar neutrino spectroscopy over the whole energy range. 
Solar neutrinos will continue to enrich our
knowledge with the existing and future experiments.


\begin{thebibliography}{99}

\bibitem{cl1}R. Davis, D.S.Harmer and K.C. Hoffman, Phys. Rev.
.Lett {\bf 20}, 1205 (1968).

\bibitem{bruno}B. Pontecorvo, JETP {\bf{6}}, 429 (1958); Z. Maki, M.
Nakagawa and S. Sakata, Prog. Theor. Phys. {\bf{28}}, 870 (1962).

\bibitem{kam1}K.S. Hirata {\em et al.}, Phys. Rev. {\bf D44}, 2241
(1991).

\bibitem{gallex1}P. Anselman {\em et al.},Phys.Lett {\bf B314},
284 (1993).

\bibitem{sage1}A.I. Abazov {\em et al.}, Phys. Rev. Lett {\bf
B67}, 3332 (1991).

\bibitem{sksolar}Y. Fukuda \etal (The Super-Kamiokande collaboration),
Phys. Rev. Lett. {\bf 81}, 1158 (1998); erratum {\bf 81}, 4279
(1998).

\bibitem{snocc} The SNO Collaboration (Q.R. Ahmad {\it et al.}),
Phys. Rev. Lett. {\bf 87}, 071301 (2001)

\bibitem{snonc}
The SNO Collaboration (Q.R. Ahmad {\it et al.}), (submitted to
Phys. Rev. Lett.), nucl-ex/0204008.

\bibitem{snodn}
The SNO Collaboration (Q.R. Ahmad {\it et al.}), (submitted to
Phys. Rev. Lett.), nucl-ex/0204009.

\bibitem{msw}L. Wolfenstein, {\em Phys. Rev.} {\bf D34}, 969 (1986);
S.P. Mikheyev and A.Yu. Smirnov,  {\em Sov. J. Nucl.
Phys.} {\bf 42(6)}, 913 (1985); {\em Nuovo Cimento} {\bf 9c}, 17 (1986).

\bibitem{kam_1}
K. Eguchi {\it et al.},
  [KamLAND Collaboration],
%``First results from KamLAND: Evidence for reactor anti-neutrino  disappearance
arXiv:hep-ex/0212021.
%%CITATION = HEP-EX 0212021;%%

\bibitem{bah}J.N. Bahcall, N.A. Bahcall, G. Shaviv, Phys. Rev. Lett.
{\bf 20}, 1209 (1968); J.N. Bahcall and Ulrich (1988); J.N.
Bahcall, M.P. Pinsonneault, Rev. Mod. Phys. {bf 64}, 885 (1992);
J.N. Bahcall and M.H. Pinsonneault,  Rev. Mod. Phys. {\bf 67}, 781
(1995).

\bibitem{bp98}
J.N. Bahcall, S. Basu, M.P. Pinsonneault, Phys. Lett. {\bf B433},
1 (1998).

\bibitem{bp00} J.N. Bahcall, S. Basu, M. Pinsonneault,
Ap. J. {\bf 555}, 990 (2001).

\bibitem{turck} S. Turck-Chi\'{e}ze,
``Review Of Solar Models And Helioseismology,''
Nucl.\ Phys.\ Proc.\ Suppl.\  {\bf 91}, 73 (2001),
S. Turck-Chi\`{e}ze and I. Lopez,  Ap. J. {\bf 408},
347 (1993); S. Turck-Chi\`{e}ze, {\it et al., ibid.} {\bf 335}, 415
(1988).

\bibitem{jnb}http://www.ias.sns.edu/~jnb.

\bibitem{kirsten}T.A. Kirsten, Rev. Mod. Phys. {\bf 71}, 1213, (1999). 

\bibitem{holm}H.D. Holmgren and R.L. Johnston, Phys. Rev. {\bf
D113}, 1556 (1959).

\bibitem{iso}J.N. Bahcall and C.A. Barnes, Phys. Lett. {\bf 12},
48 (1964).

\bibitem{cl}B.T. Cleveland {\it et al.} Astrophys. J {\bf 496}, 505 (1998).

\bibitem{sage}J.
J. N. Abdurashitov {\em et
al.}, (The SAGE collaboration), hep-ph/0204245, also see
{Phys. Rev. Lett.} {\bf 77}, 4708
(1996); Phys. Rev. {\bf C 60}, 055801 (1999).

\bibitem{gallex}
 W. Hampel {\em et al.}, (The Gallex collaboration), {Phys.
Lett.} {\bf B388}, 384 (1996); Phys. Lett. {bf B447}, 127 (1999),
Talk presented in Neutrino 2000 held at Sudbury, Canada (T.A.
Kirsten for The Gallex collaboration), Nucl. Phys. {\bf B} Proc.
Suppl. {\bf 77}, 26 (2000).

\bibitem{gno}
M. Altmann {\it et al.}, (The GNO collaboration),Phys. Lett. {\bf
B492},16 (2000); C.M. Cattadori, Nucl. Phys. {\bf B110}, Proc.
Suppl, 311 (2002).

\bibitem{kam}Y. Fukuda {\em et al.}, (The
Kamiokande collaboration), {Phys. Rev. Lett.} {\bf 77}, 1683
(1996).
%\bibitem{sk1258}Y. Fukuda {\it et al.}, Phys. Rev. Lett.
%{\bf 86}, 5651 (2001).

\bibitem{smy2002}
M.~B.~Smy, hep-ex/0202020.

\bibitem{assd}
A.~Bandyopadhyay, S.~Choubey, S.~Goswami and D.~P.~Roy,
%``Implications of the first neutral current data from SNO for solar  neutri
no oscillation,''
Phys.\ Lett.\ B {\bf 540}, 14 (2002)
[arXiv:hep-ph/0204286].
%%CITATION = HEP-PH 0204286;%%

\bibitem{kl}
%\cite{Barger:2002at}
%\bibitem{Barger:2002at}
V.~Barger and D.~Marfatia,
%``KamLAND and solar neutrino data eliminate the LOW solution,''
arXiv:hep-ph/0212126;
%%CITATION = HEP-PH 0212126;%%
%
%\cite{Fogli:2002au}
%\bibitem{Fogli:2002au}
G.~L.~Fogli, E.~Lisi, A.~Marrone, D.~Montanino, A.~Palazzo and A.~M.~Rotunno,
%``Solar neutrino oscillation parameters after first KamLAND results,''
arXiv:hep-ph/0212127;
%%CITATION = HEP-PH 0212127;%%
%
%\cite{Maltoni:2002aw}
%\bibitem{Maltoni:2002aw}
M.~Maltoni, T.~Schwetz and J.~W.~Valle,
%``Combining first KamLAND results with solar neutrino data,''
arXiv:hep-ph/0212129;
%%CITATION = HEP-PH 0212129;%%
%hcall:2002ij}
%\cite{Bandyopadhyay:2002en}
\bibitem{Bandyopadhyay:2002en}
A.~Bandyopadhyay, S.~Choubey, R.~Gandhi, S.~Goswami and D.~P.~Roy,
%``The solar neutrino problem after the first results from KamLAND,''
arXiv:hep-ph/0212146.
%%CITATION = HEP-PH 0212146;%%
%\bibitem{Bahcall:2002ij}
J.~N.~Bahcall, M.~C.~Gonzalez-Garcia and C.~Pena-Garay,
%``Solar neutrinos before and after KamLAND,''
arXiv:hep-ph/0212147;
%%CITATION = HEP-PH 0212147;%%
%
%\cite{deHolanda:2002iv}
%\bibitem{deHolanda:2002iv}
P.~C.~de Holanda and A.~Y.~Smirnov,
%``LMA MSW solution of the solar neutrino problem and first KamLAND  results,''
arXiv:hep-ph/0212270.
%%CITATION = HEP-PH 0212270;%%
%
%\cite{Nunokawa:2002mq}
%\bibitem{Nunokawa:2002mq}
H.~Nunokawa, W.~J.~Teves and R.~Zukanovich Funchal,
%``Determining the oscillation parameters by solar neutrinos and KamLAND,''
arXiv:hep-ph/0212202;
%%CITATION = HEP-PH 0212202;%%
%
%\cite{Aliani:2002na}
%\bibitem{Aliani:2002na}
P.~Aliani, V.~Antonelli, M.~Picariello and E.~Torrente-Lujan,
%``Neutrino mass parameters from Kamland, SNO and other solar evidence,''
arXiv:hep-ph/0212212;
%%CITATION = HEP-PH 0212212;%%
%
%\cite{Balantekin:2003dc}
%\bibitem{Balantekin:2003dc}
A.~B.~Balantekin and H.~Yuksel,
%``Global analysis of solar neutrino and KamLAND data,''
arXiv:hep-ph/0301072;
%%CITATION = HEP-PH 0301072;%%
%
%\cite{Creminelli:2001ij}
%\bibitem{Creminelli:2001ij}
P.~Creminelli, G.~Signorelli and A.~Strumia,
%``Frequentist analyses of solar neutrino data,''
%JHEP {\bf 0105}, 052 (2001)
[arXiv:hep-ph/0102234 (v4)].
%%CITATION = HEP-PH 0102234;%%

\bibitem{borexino}For a recent reference see G. Alimonti {\it et al.}, 
[Borexino Collaboration], Astropart. Phys., {\bf 16}, 205, 
2002. 

\bibitem{li}A. Kopylov and V. Petkhov, hep-ph/0301016. 

\bibitem{scho}The talk by S. Sch\"{o}nert, 
http://neutrino2002.ph.tum.de. 

\bibitem{hellaz}A.de. Bellefon, (The HELLAZ Collaboration), Nucl. Phys. 
Proc. Suppl., {\bf 70}, 386 (1999). 

\bibitem{heron}http://www.physics.brown.edu/research/cme/heron/. 

\bibitem{munu}
C. Arpesella, C. Broggini and C. Cattadori, Astroparticle Physics, 
{\bf 4}, 333 (1996).

\bibitem{genius}H.V. Klapdor-Kleingrothaus, hep-ph/0206249 and references 
therein. 

\end{thebibliography}
\end{document}